\documentclass[11pt,a4paper]{article}
\usepackage[latin1]{inputenc}
\usepackage{amsmath}
\usepackage{amsfonts}
\usepackage{amssymb}
\usepackage{amsthm}
\usepackage[x11names, rgb]{xcolor}
\usepackage{tikz}
\usepackage{authblk}
\usetikzlibrary{snakes,arrows,shapes}

\title{Computational Asymmetry in Strategic Bayesian Networks}
\author[1]{Sebastian Benthall}
\author[1]{John Chuang}
\affil[1]{School of Information, UC Berkeley}
\affil[1]{\textit{\{sb,chuang\}@ischool.berkeley.edu}}

\begin{document}

\maketitle

\begin{abstract}
Among the strategic choices made by today's economic actors are choices about algorithms and computational resources.
Different access to computational resources may result in a kind of economic asymmetry analogous to information asymmetry.
In order to represent strategic computational choices within a game theoretic framework, 
we propose a new game specification, Strategic Bayesian Networks (SBN).
In an SBN, random variables are represented as nodes in a graph, with edges indicating probabilistic dependence.
For some nodes, players can choose conditional probability distributions as a strategic choice.
Using SBN, we present two games that demonstrate computational asymmetry.
These games are symmetric except for the computational limitations of the actors.
We show that the better computationally endowed player receives greater payoff.
\end{abstract}

In competitive arenas such as computational finance, web search and advertising, Internet security, and AI challenges,
adoption of computer programs is an important strategic choice made by economic agents.
One limiting factor to the adoption of these programs is the computational resources available for processing.
This introduces a new potential for economic asymmetry, analogous to information asymmetry: computational asymmetry.
Actors may be symmetrical in all respects except for their capacity to compute.
Intuitively, we might guess that more computationally powerful players will often be better off.
Under what conditions is this the case, and how can we model situations where computational strategies are in play?
Algorithmic game theory offers a strong foundation for analyzing this phenomenon,
but we believe a new theoretical apparatus would assist us.

In order to demonstrate the potential effect of computational asymmetry on economic agents, we develop a new kind of game specification:
a Strategic Bayesian network (SBN).
Strategic Bayesian networks are an extension of Bayesian networks, a powerful form of representing conditional probabilities over many random variables, and influence diagrams, which extend Bayes networks to represent decision problems.
SBN further extends the Bayes network model to include strategic choices by multiple actors.

In Section 1, we will define SBN and show that games in SBN form are reducible to extensive form, 
proving that other concepts from game theory like theory of Nash Equilibrium can be applied to them. 
In Section 2, we will show how SBN is able to model computational strategies in games in a way that is 
difficult for extensive form, because of the explosive expansion of the decision tree.  
In Section 3, we will use SBN to demonstrate the phenomenon of computational asymmetry in a trivial sense:
where one economic actor is unable to process all the available input data pertinent to game outcome.
In Section 4, we demonstrate a more substantive kind of computational asymmetry, where the stronger
player is able to consistently outsmart the weaker player due to the computational complexity of
the game itself.
We will conclude with implications and directions for future research in Section 5.

\section{Strategic Bayesian networks}

Bayesian Networks are widely used as a representation of probabilistic knowledge and reasoning.  
A Bayes net is a graphical model in which nodes represent random variables and edges show the conditional dependencies of nodes on each other.
The complete distribution over variables is defined by this structure and explicit probability distribution for the value of each node, conditional on the node's parents.
There are efficient ways of simulating and computing the result of these distributions \cite{aibook}.
We propose Strategic Bayesian networks as a game theoretic extension to Bayes networks.

The intuition behind Strategic Bayes networks is that we can make these probabilistic graphical models into games by allowing actors to choose the conditional probability function at some of a graph's nodes, and assigning the actors payoffs based on the values of other nodes.

We can think of this game having two stages.
In the first stage, players look at the interconnected events in the game domain and 
determine how to ``program'' the nodes that are under their control.
When all strategically controlled nodes are programmed, the SBN is a fully specified Bayesian Network, with the 
augmentation that some nodes correspond to player utility.
In the second stage, the game ``happens'', and the events are sampled according to the Bayes network distribution.
The players reap benefits or penalties depending on the value of the payoff nodes.
To win the game, the players need to program the SBN in a way that will maximize their expected utilities.

Though graphical, SBN's are significantly different from \emph{graphical games}.
Graphical games are used to compactly represent games with very large numbers of players.
A node represents a player, whose payoff depends only the actions of itself and the players
from which it has incoming edges.
Kearns \cite{mkearns}, comparing graphical games with graphical models for
probabilistic inference including (non-strategic) Bayesian networks,
writes that ``in probabilistic inference, interactions are stochastic, whereas in graphical games
they are strategic (best response).''
We propose that Strategic Bayes Networks can succinctly bridge probabilistic inference
and game theory to model situations with both stochastic and strategic interactions.

\subsection{Definition of a Strategic Bayesian network}

Define a Strategic Bayesian network (SBN) over a directed acyclic graph, $G = (V,E)$, where $V$ are nodes and $E$ are edges.
In a standard Bayes network, each node represents a random variable whose probability function takes the possible values of its parents as input.
For an SBN, one extends this model with the following:
\begin{itemize}
\item $n$ players
\item The nodes $V$ partitioned into three subsets:
\begin{itemize}
\item $C$, a set of Chance nodes
\item $R$, a set of Strategic nodes,
\item $\Pi$, a set of Payoff nodes, one per player
\end{itemize}
\item For every Chance node $c \in C$, a probability function $f_c$ that defines, for every possible value of that $c$'s parents, a probability distribution over a set of possible values $D_c$
\item For every Strategic node $r \in R$, a player associate with that node $q(r)$ and a set of possible actions, $A_r$.  Each member of $A_r$ must be a function from the possible values of $r$'s parent nodes to a probability distribution over the nodes's possible values, $D_r$.
\footnote{A simpler variation of SBN would constrain the possible actions $a \in A_r$ to functions that produce deterministic outcomes.  This constraint would not result in loss of generality since one could construct the game graph with a single-parent Chance node as the child of every Strategic node.  However, this would break the correspondence between a fully programmed SBN and classic Bayes Net.
We also leverage the probabilistic nature of the strategic nodes in one of our proofs below to simulate the choice
of a mixed strategy in a strategic subgame.} 
\item For every node $\pi \in \Pi$, an associated player $q(\pi)$ and a function $f_\pi$ from the possible values of its parents to a probability distribution over real numbers $\Re$, which is the utility awarded to the player upon playing the game.
To simplify graphical representation, in games below we elide these payoff nodes into a single node,
whose values can be thought of as tuples of real numbers that correspond to each players' payoff in order.
\end{itemize} 

When presenting an SBN graph visually, we have adopted the notational convention of denoting Chance nodes with a single elliptical outline, 
Strategic nodes with a double elliptical outline, and Payoff nodes with a rectangular outline.

When players play a SBN game, the entire specification of the SBN is made available to them.
Then, without coordination or observation, for each strategic node $r \in R$ the player $q(r)$ picks an action $a_r \in A_r$.
Players choose their action rationally in an effort to maximize their expected utility from the payoff nodes $\Pi$.
Once the actions have been chosen, all the random variables are sampled once according to the probability distributions $f_i$ and $a_i$.
The nodes are sampled in depth order.
In this way, first the values of nodes with no parents are determined (their probability functions take no inputs), and each remaining node can take its inputs from values already sampled.
When all nodes are determined, payoffs are awarded to the players.

\subsection{Reduction of SBN to Extensive Form Game}

In order to show that SBN games are the kind of game to which we can apply game theoretic notions such as best response
and Nash equilibrium, we here sketch the reduction from an SBN to an extensive form game.

To perform the reduction, construct a game tree starting with each of the strategic nodes $r \in R$, with the same number of players as the SBN.

For the root of the tree, pick a $\rho \in R$ and make a corresponding root node $v_\rho$.
Let $v_\rho$ be a node of player $q(\rho)$ with a branch for each element of $A_\rho$.

For each remaining $r \in R$, put a node that belongs to player $q(r)$ at the end of the branches of the preceding nodes.
Give them each a branch for every action in $A_r$.

At this point, the constructed game tree will have several tiers of non-Chance nodes.
Each tier will correspond to a single Strategic node in the SBN, and all the nodes in a tier will have the
same player.
Each leaf of this tree $w$ will correspond to a set of choices, one for each Strategic node in the SBN.
In other words, each leaf ``fully programs'' the original SBN.

Next comes the tricky part: putting these nodes into the appropriate information sets.
Recall that in an extensive form game, an information set is a set of nodes between which a player cannot discriminate at the moment of their choice.
In an SBN, each player knows their own action choices, but not other players.
In the extensive form reduction of an SBN, players need \emph{perfect recall} \cite{stengel} but have no knowledge of other players' moves.
To accomplish this in the reduction, apply the following rule:
for every non-chance node in the tree corresponding to a Strategic node $r$ in the SBN,
assign it to an information set with every other tree node corresponding to $r$ that
\emph{shares in its path to the tree root any and all actions played at other nodes by player $q(r)$}.
Thus, in the `sequential' extensive form of an SBN, a player's choice of action is made with only the knowledge of that players' previous actions.

For each leaf $w$, unfold this Bayesian network into a subtree of (extensive form) Chance nodes, in depth order.
For each node in the Bayesian network $u$, put a node at each leaf of the graph that branches for all values in $D_u$ with probabilities assigned by the distribution returned by $f_u$ or $a_u$ (whichever is appropriate) when given the input values represented by the the path from the node to the subtree root.

When a layer of payoff nodes are introduced in this way, branch on a Chance node as usual but label that branch with the payoff awarded for the corresponding player according to $f_\pi$.

When all of the Bayesian network nodes have been used to generate game tree nodes in this way, assign to each leaf the n-tuple of payoffs for each player based on the labels on the path from it to the game tree root.

This concludes the construction of the extensive form game from an SBN.  The number of nodes in this tree will be equal to $1 + \prod_{u \in V} D_u$, not including the payoff leaves.  This number may be infinite.

\section{Computational strategies}

As we have discussed, we believe that an advantage of the Strategic Bayesian network 
formalism is that it can model games in which a player's choice of moves is subject to 
complexity constraints.
One way to do this is to restrict the choice of actions at a players' strategic node to a
complexity class.
We believe that this technique has the expressive power to represent computational games in 
a way that is difficult to do with extensive form games alone.

\subsection{A simple example}

Consider playing the following single player game, \textsc{NoCount}: 
you receive a string of random bits of length $n$, each generated from an even coin toss.
The variable $n$ has been generated from an exponential distribution with parameter $\lambda$.
You are allowed to write a program that takes the string as input and returns an integer $g$ between 0 and n.
If $g$ is equal to the sum of all the 1's in the string, you get a prize.
Otherwise, you get nothing.
The catch is that your program must be an algorithm that runs in constant time in the worst case.

We can specify this game using an SBN with the following graphical structure:

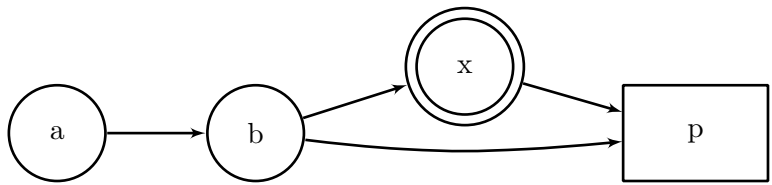
\begin{tikzpicture}[>=latex',line join=bevel,]
  \pgfsetlinewidth{1bp}
\pgfsetcolor{black}
  \draw [->] (37.673bp,18bp) .. controls (45.736bp,18bp) and (55.352bp,18bp)  .. (74.333bp,18bp);
  \draw [->] (110.7bp,23.673bp) .. controls (119.33bp,26.439bp) and (129.97bp,29.848bp)  .. (149.48bp,36.102bp);
  \draw [->] (192.7bp,36.691bp) .. controls (200.96bp,34.291bp) and (210.62bp,31.482bp)  .. (229.75bp,25.921bp);
  \draw [->] (111.54bp,15.469bp) .. controls (122.19bp,14.144bp) and (135.82bp,12.662bp)  .. (148bp,12bp) .. controls (168.41bp,10.89bp) and (173.58bp,11.03bp)  .. (194bp,12bp) .. controls (202.35bp,12.397bp) and (211.3bp,13.087bp)  .. (229.75bp,14.835bp);
\begin{scope}
  \definecolor{strokecol}{rgb}{0.0,0.0,0.0};
  \pgfsetstrokecolor{strokecol}
  \draw (19bp,18bp) ellipse (18bp and 18bp);
  \draw (19bp,18bp) node {a};
\end{scope}
\begin{scope}
  \definecolor{strokecol}{rgb}{0.0,0.0,0.0};
  \pgfsetstrokecolor{strokecol}
  \draw (171bp,43bp) ellipse (18bp and 18bp);
  \draw (171bp,43bp) ellipse (22bp and 22bp);
  \draw (171bp,43bp) node {x};
\end{scope}
\begin{scope}
  \definecolor{strokecol}{rgb}{0.0,0.0,0.0};
  \pgfsetstrokecolor{strokecol}
  \draw (93bp,18bp) ellipse (18bp and 18bp);
  \draw (93bp,18bp) node {b};
\end{scope}
\begin{scope}
  \definecolor{strokecol}{rgb}{0.0,0.0,0.0};
  \pgfsetstrokecolor{strokecol}
  \draw (284bp,36bp) -- (230bp,36bp) -- (230bp,0bp) -- (284bp,0bp) -- cycle;
  \draw (257bp,18bp) node {p};
\end{scope}
\end{tikzpicture}

Here, $\lambda$ is a variable exogenous to the model.
The function $f_a$ is the exponential distribution (or, better, a distribution with an undefined mean) with parameter $\lambda$, rounded up.  
The function $f_b$ is the function that takes an integer $n$ and returns $s$, a string of 0's and 1's according to a fair coin toss.
I.e., it is the result of a Bernoulli process of length $n$ with $p=.5$.  $A_x = O(1)$.  
$f_\pi$ takes two inputs, $g$ and $s$ and returns a deterministic distribution, either $1$ if $g=\sum s$ and $0$ otherwise.

We make two observations.
First, the complexity restriction is a frustration for the player of \textsc{NoCount}.  
Without this restriction, the player could program node $x$ to simply count the number of 1's in $s$ and return the answer.
But as this is a linear time function, this is not a possible choice.
Rather, the player must do the best she can in constant time.
We raise as a conjecture that the player's best to choice for $a_x$ is the function that ignores its input and returns $round(\frac{1}{2\lambda})$.
We leave the proof or refutation of this claim as an open problem.

Second, we note that this SBN representation is quite compact compared to an equivalent extensive form game.
Shoehorning the problem into a formal game tree definition would require a separate derivation from the complexity class $O(1)$ at almost every level of the formalism: the predecessor function, the partitioning into information sets, the action set, and the probability function determining the behavior of the nature nodes (recall that there is a separate nature node for every $O(1)$ function.
We leave this exercise of specifying \textsc{NoCount} as an extensive form game as another open problem.

\section{Trivial Computational Asymmetry}

We have shown how to use Strategic Bayes Networks to model games with complexity constraints.
We now want to explore games with computational asymmetry, which we believe is a new concept in economics.  
The intuition behind computational asymmetry is that when players in an otherwise symmetric game have different 
capacities resources at their disposal, this condition can provide an advantage to the better computationally endowed player.

Consider a new game, \textsc{TwoPlayerNoCount}, with this SBN structure:

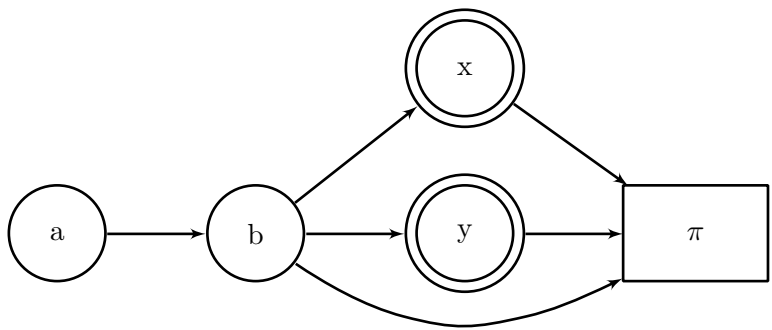
\begin{tikzpicture}[>=latex',line join=bevel,]
  \pgfsetlinewidth{1bp}
\pgfsetcolor{black}
  \draw [->] (111.88bp,34.753bp) .. controls (119.79bp,34.753bp) and (129.22bp,34.753bp)  .. (148.41bp,34.753bp);
  \draw [->] (107.67bp,46.415bp) .. controls (118.31bp,54.875bp) and (132.98bp,66.534bp)  .. (153.39bp,82.756bp);
  \draw [->] (189.24bp,83.604bp) .. controls (199.24bp,76.396bp) and (211.94bp,67.235bp)  .. (231.94bp,52.821bp);
  \draw [->] (193.6bp,34.753bp) .. controls (201.61bp,34.753bp) and (210.85bp,34.753bp)  .. (229.97bp,34.753bp);
  \draw [->] (37.673bp,34.753bp) .. controls (45.736bp,34.753bp) and (55.352bp,34.753bp)  .. (74.333bp,34.753bp);
  \draw [->] (107.96bp,23.513bp) .. controls (118.53bp,16.301bp) and (133.37bp,7.5923bp)  .. (148bp,3.753bp) .. controls (167.77bp,-1.4361bp) and (174.09bp,-0.88074bp)  .. (194bp,3.753bp) .. controls (202.97bp,5.8407bp) and (212.14bp,9.3968bp)  .. (229.8bp,17.984bp);
\begin{scope}
  \definecolor{strokecol}{rgb}{0.0,0.0,0.0};
  \pgfsetstrokecolor{strokecol}
  \draw (19bp,35bp) ellipse (18bp and 18bp);
  \draw (19bp,34.753bp) node {a};
\end{scope}
\begin{scope}
  \definecolor{strokecol}{rgb}{0.0,0.0,0.0};
  \pgfsetstrokecolor{strokecol}
  \draw (171bp,97bp) ellipse (18bp and 18bp);
  \draw (171bp,97bp) ellipse (22bp and 22bp);
  \draw (171bp,96.753bp) node {x};
\end{scope}
\begin{scope}
  \definecolor{strokecol}{rgb}{0.0,0.0,0.0};
  \pgfsetstrokecolor{strokecol}
  \draw (93bp,35bp) ellipse (18bp and 18bp);
  \draw (93bp,34.753bp) node {b};
\end{scope}
\begin{scope}
  \definecolor{strokecol}{rgb}{0.0,0.0,0.0};
  \pgfsetstrokecolor{strokecol}
  \draw (171bp,35bp) ellipse (18bp and 18bp);
  \draw (171bp,35bp) ellipse (22bp and 22bp);
  \draw (171bp,34.753bp) node {y};
\end{scope}
\begin{scope}
  \definecolor{strokecol}{rgb}{0.0,0.0,0.0};
  \pgfsetstrokecolor{strokecol}
  \draw (284bp,53bp) -- (230bp,53bp) -- (230bp,17bp) -- (284bp,17bp) -- cycle;
  \draw (257bp,34.753bp) node {$\pi$};
\end{scope}
\end{tikzpicture}

In this game, $a$ and $b$ are defined as before.
But $x$ and $y$ are controlled by different players.
The values of $b$, $x$, and $y$ all go to to determine the payoffs at $\pi$.
If both guesses are correct (i.e. $x = y = sum(b)$), the two players split the prize.
If neither guesses correctly, neither gets the prize.
If only one guesses correctly, the player that guessed correctly wins the prize.

To introduce the idea of \textbf{computational asymmetry}, we will give the player controlled nodes different action sets.
Let $A_x = O(1)$.
But let $A_y = O(n)$.

$y$'s player can guarantee greater expected utility in this game because summing the 1's in the string from $b$ is within their action set, whereas $x$'s player has to find some constant time approximation.
In this otherwise symmetric game, the player that controls $y$ has an advantage that is due to a difference in computational power.

\subsection{Some early conclusions}

The \textsc{TwoPlayerNoCount} example suggests several concepts that could come into play in other situations with computational asymmetry.

First, note that the game structure faces the players with a decision problem with a certain computational complexity--
in this case, linear complexity.
If both players were able to pick strategies of this complexity or greater, then asymmetry would have no effect on expected utility.
Computational asymmetry surfaces when one or more players is insufficiently equipped to guarantee them the maximum possible utility in the game.

Second, note that the computational power of one player in this example negatively effects the other player.
Because $y$'s player \emph{always} guesses the correct value, $x$'s player will never get the full prize.
Meanwhile, $y$'s player can capitalize on the $x$ players' lack of computational power by claiming the full prize whenever the latter guesses incorrectly.

Third, the $x$ players' disadvantage is due to the fact that $x$ is not able to choose a strategy 
that uses all the information available to it.
More specifically, though $x$'s player has perfect information about the game itself, 
the function at $x$ loses information provided to it by its conditioning variables.

This last point explains a trivial way that computational asymmetry can effect game outcome.
More interesting and analytically challenging cases involve situations where strategic nodes 
are computationally strong enough to read their entire "input" from conditioning variables, 
but are not strong enough to "solve" for the function that would maximize their player's payoff.

To illustrate such a case, we will introduce one more game.

\section{Substantive Computational Assymetry}

\textsc{LetsPlay} is a game defined by an SBN with two players.
It has the structure shown below, with the payoff nodes elided for simplicity.

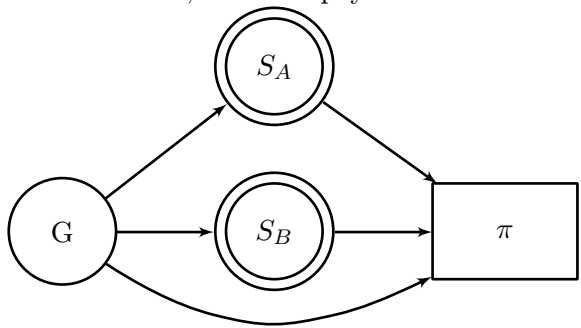
\begin{tikzpicture}[>=latex',line join=bevel,]
  \pgfsetlinewidth{1bp}
\pgfsetcolor{black}
  \draw [->] (40.344bp,34.715bp) .. controls (48.296bp,34.715bp) and (57.615bp,34.715bp)  .. (76.499bp,34.715bp);
  \draw [->] (117.24bp,83.566bp) .. controls (127.24bp,76.358bp) and (139.94bp,67.197bp)  .. (159.94bp,52.783bp);
  \draw [->] (36.21bp,22.842bp) .. controls (46.915bp,15.758bp) and (61.583bp,7.4394bp)  .. (76bp,3.7151bp) .. controls (95.795bp,-1.3982bp) and (102.09bp,-0.91869bp)  .. (122bp,3.7151bp) .. controls (130.97bp,5.8028bp) and (140.14bp,9.3589bp)  .. (157.8bp,17.946bp);
  \draw [->] (121.6bp,34.715bp) .. controls (129.61bp,34.715bp) and (138.85bp,34.715bp)  .. (157.97bp,34.715bp);
  \draw [->] (35.987bp,47.262bp) .. controls (46.683bp,55.657bp) and (60.997bp,66.89bp)  .. (81.388bp,82.893bp);
\begin{scope}
  \definecolor{strokecol}{rgb}{0.0,0.0,0.0};
  \pgfsetstrokecolor{strokecol}
  \draw (99bp,97bp) ellipse (18bp and 18bp);
  \draw (99bp,97bp) ellipse (22bp and 22bp);
  \draw (99bp,96.715bp) node {$S_A$};
\end{scope}
\begin{scope}
  \definecolor{strokecol}{rgb}{0.0,0.0,0.0};
  \pgfsetstrokecolor{strokecol}
  \draw (212bp,53bp) -- (158bp,53bp) -- (158bp,17bp) -- (212bp,17bp) -- cycle;
  \draw (185bp,34.715bp) node {$\pi$};
\end{scope}
\begin{scope}
  \definecolor{strokecol}{rgb}{0.0,0.0,0.0};
  \pgfsetstrokecolor{strokecol}
  \draw (99bp,35bp) ellipse (18bp and 18bp);
  \draw (99bp,35bp) ellipse (22bp and 22bp);
  \draw (99bp,34.715bp) node {$S_B$};
\end{scope}
\begin{scope}
  \definecolor{strokecol}{rgb}{0.0,0.0,0.0};
  \pgfsetstrokecolor{strokecol}
  \draw (20bp,35bp) ellipse (20bp and 20bp);
  \draw (20bp,34.715bp) node {G};
\end{scope}
\end{tikzpicture}

$G$ is a Chance node.
Its range of possible values $D_G$ is all possible symmetric zero sum games with payoff matrix sized $n \times n$.
For our purposes, $n$ may be drawn from any distribution that assigns positive mass to all positive integers, such as an exponential distribution, and payoff matrix values are populated from a normal distribution rounded to some decimal place, subject to the symmetry and zero sum constraints.
This will guarantee that for roughly any symmetric game $g$ there is some probability mass $P(G = g)$. 

$S_a$ and $S_b$ are strategic nodes for players A and B, respectively.
For both nodes, their range of possible values is a positive integer index into the matrix representing the subgame--or, a pure strategy
for the subgame $g$.
The action sets $A_S$ are both subsets of the functions that map from \textit{game matrices} to \textit{mixed strategies in those games}.
Recall that mixed strategies are themselves distributions over pure strategies.
So these strategic nodes fulfill the SBN requirement that action sets must be comprised of functions from node inputs to probability
distributions over possible node values.

Consider for a moment some well-established results in algorithmic game theory.
Nash proved that symmetric games always have a symmetric Nash equilibrium in which 
all players play identical mixed strategies.\cite{nash}
Papadimitriou \cite{cpap} has shown that the problem of finding a symmetric Nash
Equilibrium of a symmetric game is PPAD-hard, and notes that the problem
of finding \emph{symmetric or non-symmetric} Nash equilibrium to symmetric games
is possibly easier.
It is well-known that solving for Nash equilibria of zero-sum games
in normal form is reducible to a linear programming problem.\cite{etardos}

This proof depends on the complexity of two problems on the domain of
normal form games that are both symmetric and zero sum.
The first, which we will call \textsc{zero sum symmetric nash} is the problem of finding a 
strategy $m$ such that if played by both players, there is a symmetric Nash equilibrium.

The second problem, which we will call the \textsc{responder} problem,
is: given a symmetric zero sum game in normal form $g$, find a mixed strategy
$m$ such that there exists a symmetric Nash equilibrium strategy $n$ for $g$
and $m$ is a best response to $n$.

We are not aware of theorems that determine the complexity of these problems,
but our proof depends only on the \emph{existence} of their worst-case
complexity bounds.
Note that solving \textsc{zero sum symmetric nash} solves \textsc{responder}, so
responder is at most as hard as \textsc{zero sum symmetric nash}.

To introduce computational asymmetry into \textsc{LetsPlay}, we pick a universal computing language 
and a machine implementation for the purpose of evaluating function complexity.
Next, we define as $O(w(n))$ the tightest worst case performance bound for an algorithm 
that can solve \textsc{zero sum symmetric nash}.
If this is a polynomial complexity problem, for example, $w(n)$ might be $n^k$ for some constant $k$.
We restrict $S_a$ to functions that are within this complexity bound.
We then restrict $S_b$ to a strictly lesser bound.
Let $O(x(n))$ be the tightest worst case performance bound for any algorithms that can 
solve \textsc{responder}.
Define $y$ to be any function that is strictly less than $x$ such that
$w(n) \geq x(n) > y(n)$ for all $n$.
$S_b$ is limited to functions in $O(y(n))$.
The upshot of this is that Player A will be able to pick strategy $N$, which computes a symmetric Nash Equilibrium
strategy given any game $g$, and Player B will not be able to use a strategy that reliable computes best responses
to $N(g)$.

The payoff nodes in $\pi$ simply award players according their pure strategies and the payoff matrix provided by $G$.

\subsection{Proof of computational asymmetry effect on utility}

Player A can guarantee that they ``win'' this game by playing strategy $N$.
By this we mean that Player A can guarantee that they have positive expected payoff and
Player B has negative expected payoff, despite the game being perfectly symmetrical
except for the computational limits on the players' strategies.

Why is this the case?
Suppose Player A plays strategy $N$, meaning that for any payoff matrix $g$,
$S_a(g) = N(g)$ and $N(g)$ is a symmetric Nash Equilibrium strategy for $g$.
For at least some games $g$, $S_b(g) \neq N(g)$, because otherwise
$S_b = N$ and this is not allowed given Player B's computational range.

For the games where $S_b(g) = N(g)$, the expected payoff of both players is
0, since the strategies are identical and the game is symmetric and zero sum.

For the cases where $S_b(g) \neq N(g)$, we consider whether $S_b(g)$ is a best
response to $N(g)$.
If $S_b(g)$ is a best response to $N(g)$, then for all strategies $m$,
the payoff of playing $m$ against $N(g)$, $u(m,N(g))$, must be no greater
than $u(S_b(g),N(g))$.
However, $N(g)$ is a best response to itself, so $u(N(g),N(g)) \geq u(S_b(g),N(g))$.
Hence, $u(S_b(g),N(g)) = u(N(g),N(g)) = 0$ for these cases

However, Player B is not able to reliably solve the \textsc{responder} problem
because of its computational range.
Therefor, there must be some $g$
such that $S_b(g) \neq N(g)$ and $S_b(g)$ \emph{is not a best response to} $N(g)$.
In these cases, Player B's expected payoff in playing $S_b(g)$ against
$N(g)$ must be negative, because $u(S_b(g),N(g)) < u(N(g),N(g)) = 0$.
Since $g$ is a payoff matrix for a zero sum game,  Player A's payoff in
these cases must be positive.

Consider \textsc{LetsPlay} as a whole.
If Player A picks $S_a = N$, then whatever Player B chooses for $S_b$, then
the expected payoff for Player A must be positive, because there will be
at least one subgame $g$ with positive probability for which A gets positive
expected payoff, and the other games will have tied payoffs at 0.
So, Player A can guarantee that they win at \textsc{LetsPlay}
purely in virtue of the game's computational asymmetry.

\subsection{Discussion and Potential Variations}

We have presented \textsc{LetsPlay} as a simple example of how computational asymmetry
might affect economic game outcomes.
Unlike \textsc{TwoPlayerNoCount}, both players are able to ``read'' the complete ``input''
provided to them by the games' chance conditions.
However, nevertheless the computationally stronger player can guarantee their advantage.

Our argument has a weakness: if the worst-case complexity bound for \textsc{responder}
is $O(n^2)$, the same complexity as reading
the game payoff matrix, the computational asymmetry in \textsc{LetsPlay} is trivial as 
in the previous section and not what we call ``substantive''.

We have not shown whether $N$ is Player A's best available strategy, or what Player B's
best strategy is.
We have also only been able to show that Player A can eek out \emph{some} advantage,
with no regard for size or significance.
We also acknowledge that the players' computational ranges are contrived.
We hope that this result is robust enough to be suggestive of more significant results
to be discovered by brighter minds.

Several variations of \textsc{LetsPlay} have so far eluded our analysis.
We offer as a conjecture that the effect of computational asymmetry can 
be proven when one or both of the symmetry and zero-sum conditions are lifted
from the $G$ node.

\section{Discussion}

The use of abstract complexity classes in the foregoing analysis raises doubts about its applicability 
to ``real world'' problems, since these classes are defined in terms of unbounded constants.
A more compelling demonstration of computational asymmetry would involve finite bounds on computational complexity.
Future work could apply circuit complexity classes to SBN to achieve this.

We anticipate other uses of SBN beyond the proofs provided here.  
Bayesian networks are a powerful way to represent any stochastic domain with discrete events.
SBN's allow these networks to represent domains where strategic agency is a factor.
For example, SBN's might be used to model information flow within an organization.
In a principal-agent scenario, an agent may observe (be a child of) events and then file a report about them to the principal.
A principal might choose to investigate the event, reward, or reprimand the agent based on the contents of the report.
There may be several Nash equilibria of reporting and responding strategies identifiable in an SBN model of this situation.

While we have provided a formal proof using the SBN construct in this paper, 
SBN's may prove to be a flexible basis for implementing  and testing game theoretic simulations.
Recall that algorithms for computing over Bayes Networks are well-known.
Especially with more restricted action sets at strategic nodes, SBN's may provide a way to rapidly simulate strategic choices in a stochastic domain and converge on optimal or equilibrium strategies.

Overall, we hope that our mathematical analysis of computational asymmetry using Strategic Bayes Networks will pique the curiosity of
the algorithmic game theory community.
While the concept of computational asymmetry is intuitively compelling, as far as we know there have been
no attempts to model such situations game-theoretically and in a way informed by complexity theory.
We hope others can do more with the tools and ideas sketched here.

\end{document}